# Element-specific and high-bandwidth ferromagnetic resonance spectroscopy with a coherent, extreme ultraviolet (EUV) source

Michael Tanksalvala[1], Anthony Kos[1], Jacob Wisser[1], Scott Diddams[2,3,4], Hans T. Nembach[4,5], and Justin M. Shaw[1]

[1]Applied Physics Division, National Institute of Standards and Technology, Boulder, CO 80305
[2]Time & Frequency Division, National Institute of Standards and Technology, Boulder, CO 80305
[3]Electrical, Computer & Energy Engineering, University of Colorado, Boulder, Colorado 80309, USA
[4]Department of Physics, University of Colorado, Boulder, Colorado 80309, USA
[5]Associate, Physical Measurement Laboratory, National Institute of Standards and Technology, Boulder, Colorado 80305, USA

We applied a tabletop, ultrafast, high-harmonic generation (HHG) source to measure the element-specific ferromagnetic resonance (FMR) in ultra-thin magnetic alloys and multilayers on an opaque Si substrate. We demonstrate a continuous wave bandwidth of 62 GHz, with promise to extend to 100 GHz or higher. This laboratory-scale instrument detects the FMR using ultrafast, extreme ultraviolet (EUV) light, with photon energies spanning the M-edges of most relevant magnetic elements. An RF frequency comb generator is used to produce a microwave excitation that is intrinsically synchronized to the EUV pulses with a timing jitter of 1.1 ps or better. We apply this system to measure the dynamics in a multilayer system as well as Ni-Fe and Co-Fe alloys. Since this instrument operates in reflection-mode, it is a milestone toward measuring and imaging the dynamics of the magnetic state and spin transport of active devices on arbitrary and opaque substrates. The higher bandwidth also enables measurements of materials with high magnetic anisotropy, as well as ferrimagnets, antiferromagnets, and short-wavelength (high wavevector) spinwaves in nanostructures or nanodevices. Furthermore, the coherence and short wavelength of the EUV will enable extending these studies using dynamic nanoscale lensless imaging techniques such as coherent diffractive imaging, ptychography, and holography.

**Introduction**

Magnetic multilayer and device structures are the fundamental building blocks of data storage and magnetic random access memories (MRAM), as well as emerging spintronic, magnonic and spin-orbitronic devices.[1–5] As these technologies pursue faster operational speeds and push further into the deep nanoscale, new demands are placed on nanoscale characterization to evaluate their performance and elucidate the underlying physics and materials science that drives these technologies. This is especially true for dynamic phenomena, which have more limitations on the tools that are available for such studies. Ferromagnetic resonance spectroscopy (FMR) is the fundamental tool for measuring these phenomena; by driving spin precession at around the Larmor frequency and fitting the precession amplitude and phase to material-dependent terms in the Landau-Lifshitz equation, FMR provides the ability to measure magnetic properties and dynamics in thin-film systems and devices. The spin precession is usually measured as a change in the microwaves absorbed, transmitted or reflected from the sample[6,7] or through transport phenomena in fully fabricated nanodevices.[8] These approaches average over the thickness of the sample; therefore, in more complex heterostructures FMR elucidates sample-averaged properties, and measuring magnetic dynamics at individual layers or interfaces becomes difficult.

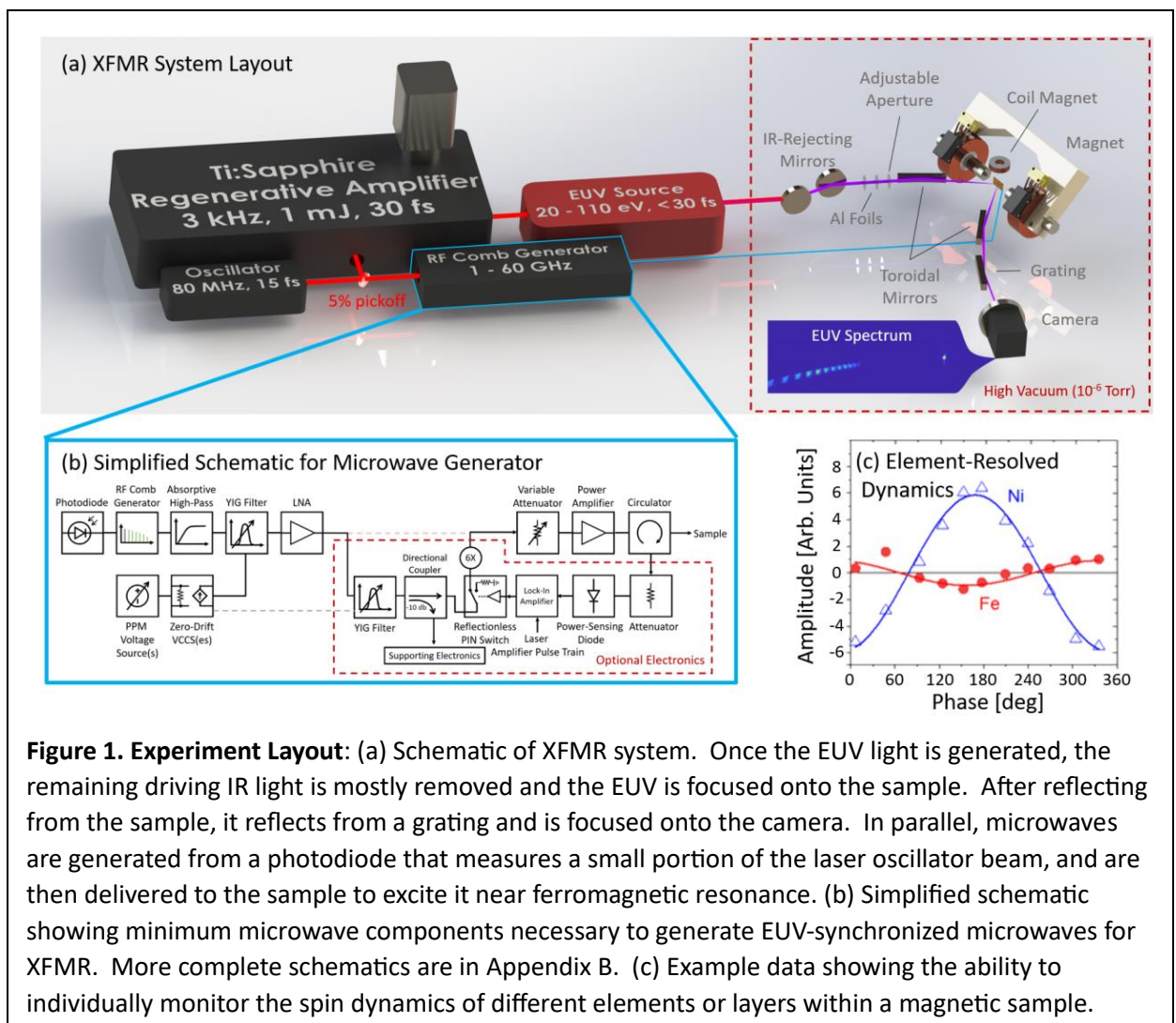

**Figure 1. Experiment Layout**: (a) Schematic of XFMR system. Once the EUV light is generated, the remaining driving IR light is mostly removed and the EUV is focused onto the sample. After reflecting from the sample, it reflects from a grating and is focused onto the camera. In parallel, microwaves are generated from a photodiode that measures a small portion of the laser oscillator beam, and are then delivered to the sample to excite it near ferromagnetic resonance. (b) Simplified schematic showing minimum microwave components necessary to generate EUV-synchronized microwaves for XFMR. More complete schematics are in Appendix B. (c) Example data showing the ability to individually monitor the spin dynamics of different elements or layers within a magnetic sample.

For over a decade, x-ray detected ferromagnetic resonance spectroscopy (XFMR) has served as an enhancement to FMR by detecting the spin precession with an element-specific probe (x-rays or extreme ultraviolet light). Higher-energy photons in the extreme ultraviolet (EUV) or x-ray — as opposed to optical or microwave wavelength — provide element-specific information through direct access to atomic core level transitions. In this technique, x-ray magnetic circular dichroism (XMCD) and FMR are combined by exciting the sample with a radio frequency (RF) or microwave source that is generated directly from the master clock of the synchrotron, and is therefore synchronized to the x-ray pulses.[9,10] The element-specific response of the sample is recorded by tuning the x-ray energy to a specific atomic core transition and measuring XMCD spectra at various time delays. The ability to measure element-resolved FMR has been important for measuring magnetic systems, since it gives the ability to decouple the magnetic dynamics of different layers or sublattices within a sample if the magnetic layers contain different magnetic elements.[11–16] Such studies have uncovered the details of phenomena such as interlayer exchange[10,17,18] and the presence of spin currents in nonmagnetic layers that can be used to identify contributions to spin-pumping.[19–23] Furthermore, XFMR has been integrated into a scanning transmission x-ray microscopy (STXM) modality to provide such dynamic measurement with nanoscale resolution.[24]

One technical limitation of synchrotron-based, time-resolved measurements is the maximum frequency that can be achieved.  This limitation primarily comes from the pulse width of the electron bunches, which typically ranges from 50–100 ps and limits measurement frequencies to only a few GHz.  The pulse width can be shortened by using low-alpha mode of operation, which reduces the pulse width to about 7–10 ps.[25,26]  This mode of operation has been used to demonstrate XFMR bandwidths up to approximately 11 GHz;[10,27,28] however, such bandwidths are of limited utility for many technological materials that have high magnetic anisotropy, short-wavelength modes that result from confinement, and/or many materials with opposing magnetic moments such as antiferromagnets.[29–33] Furthermore, extension to many non-magnetic phenomena, such as the measurement of phonon modes, also requires higher frequencies.[34]  While pump-probe approaches at femtosecond slicing and free electron laser facilities can measure ultrafast dynamics in the time domain, they do not comprise a resonant excitation of a particular mode.  In other words, an ultrafast pulsed excitation will excite every mode and frequency in the system, which can make it challenging to image, isolate or separate a specific excitation — especially if the excitation of interest is not dominant.

Advances in tabletop light sources, such as high-harmonic generation (HHG) sources, open new opportunities to advance laboratory-based measurement capability to compliment the capability of synchrotrons.  HHG sources have been used to measure ultrafast processes including magnetization dynamics, owing to the exceptionally small pulse width approaching the attosecond regime.[35,36]  In these studies, the combination of ultrafast and element-specific capability was critical to elucidate and separate many phenomena including superdiffusive spin currents, ultrafast magnon generation, optically induced spin-transfer processes, and exchange coupling between elements in an alloy.[36–44]  These studies typically rely on an optical pump-probe approach, with short pulse durations (10 to 100 fs) and intrinsic timing synchronization of the pump and probe naturally enabling measurements with high temporal resolution.

In this paper, we report on development of the first tabletop-scale XFMR instrument (shown in Fig. 1), and demonstrate the highest-frequency XFMR to date, at 17 GHz.  This instrument leverages the femtosecond pulse duration of the EUV probe to provide an intrinsic frequency limit exceeding 10 THz. Our current configuration generates a phase-stable microwave excitation up to 60 GHz by using an RF frequency comb generator driven by a photodiode at the output of the laser oscillator.  The current frequency limitation is set by the timing jitter of the microwaves (<1.1 ps, or > 100 GHz), as well as the limited precession cone angle exhibited by most samples at higher frequencies.

An important characteristic of our approach is that there is no constraint on the substrates that can be used since measurements can be performed in either transmission or reflection geometries.  This is especially advantageous when combined with imaging devices since special fabrication on membranes or specialized substrates such as MgO for fluorescence detection are not needed.  In transmission, either magnetic linear dichroism (MLD) or magnetic circular dichroism (MCD) can be used.[45–49]  Since the shorter absorption length of EUV relative to soft-x-rays limits the thickness of the sample in which the EUV can probe (10~1000 nm generally, e.g. ~340 nm for 50–70 eV light in Si),[50] the demonstration in this paper is done in reflection and uses the transverse magneto-optic Kerr effect (T-MOKE), which is routinely employed with linearly polarized EUV sources.[36]

Finally, the short-wavelength photons that span the EUV to soft-x-ray region of the spectrum are useful for nanoscopic imaging, given the wavelength of the photons. In our case, the high coherence further enables multiple lensless imaging modalities such as coherent diffractive imaging (CDI), ptychography, and holography.[51–53] Of import is that ptychography and other CDI techniques can be used in both reflection and transmission.[51] Recent work also combined XFMR with x-ray reflectometry to resolve the dynamic response as a function of depth.[54] Since EUV reflectometry combined with ptychography has already been demonstrated with HHG sources, this work opens up new possibilities for full 3D spatial imaging of high-bandwidth dynamics in nanostructures.[55]

The rest of the paper is structured as follows: First, we outline how we generate EUV light and use it to probe magnetic moments with element-specificity. Next, we describe how we generate high-frequency microwaves that are phase-synchronized to our EUV pulses. Then, we describe how we combine these to perform XFMR measurements. Finally, we show XFMR results on three samples: permalloy (8.5 GHz), CoFe (17 GHz), and a Ni/Fe multilayer (8.5 GHz, with layer-specific dynamics).

**Experiment Methods**

An overview of the setup used to generate the phase-stable microwave excitation and EUV probe is shown in Fig. 1. It involves two parallel components that are each seeded by a Ti:sapphire laser oscillator (15 fs, ~82 MHz, ~400 mW). The first is the microwave part that generates high-frequency microwaves (up to >60 GHz) to excite the sample, and that are synchronized to the EUV probe with a controllable phase shift. The second is the optical half that ultimately focuses femtosecond-duration extreme ultraviolet (EUV) light pulses onto the sample to probe the precession of spins therein. Finally, the EUV light that reflects from the sample is collected in a grating-based spectrometer.

**Extreme Ultraviolet Light Probes Element-Specific Magnetic Moment**

We use a regenerative Ti:sapphire laser amplifier to produce 3 kHz, ~1 mJ, 35 fs pulsed light with 800 nm wavelength. We use this to perform high harmonic generation (HHG) by focusing it into a hollow-core glass capillary filled with 300–700 Torr Neon gas. This produces a spectrum of discrete harmonics with photon energies spanning 35–70 eV, thus spanning the M-edges of most magnetic elements. Moreover, these harmonics are separated by 3 eV, are pulsed at a 3 kHz rate with a time duration shorter than the driving laser (typically $\leq$ 10 fs), and have a linear polarization matching that of the driving infrared (IR) light. We use horizontally polarized light so that the beamline can be in the plane of the optical table even after reflecting from the sample around Brewster's angle for the EUV.

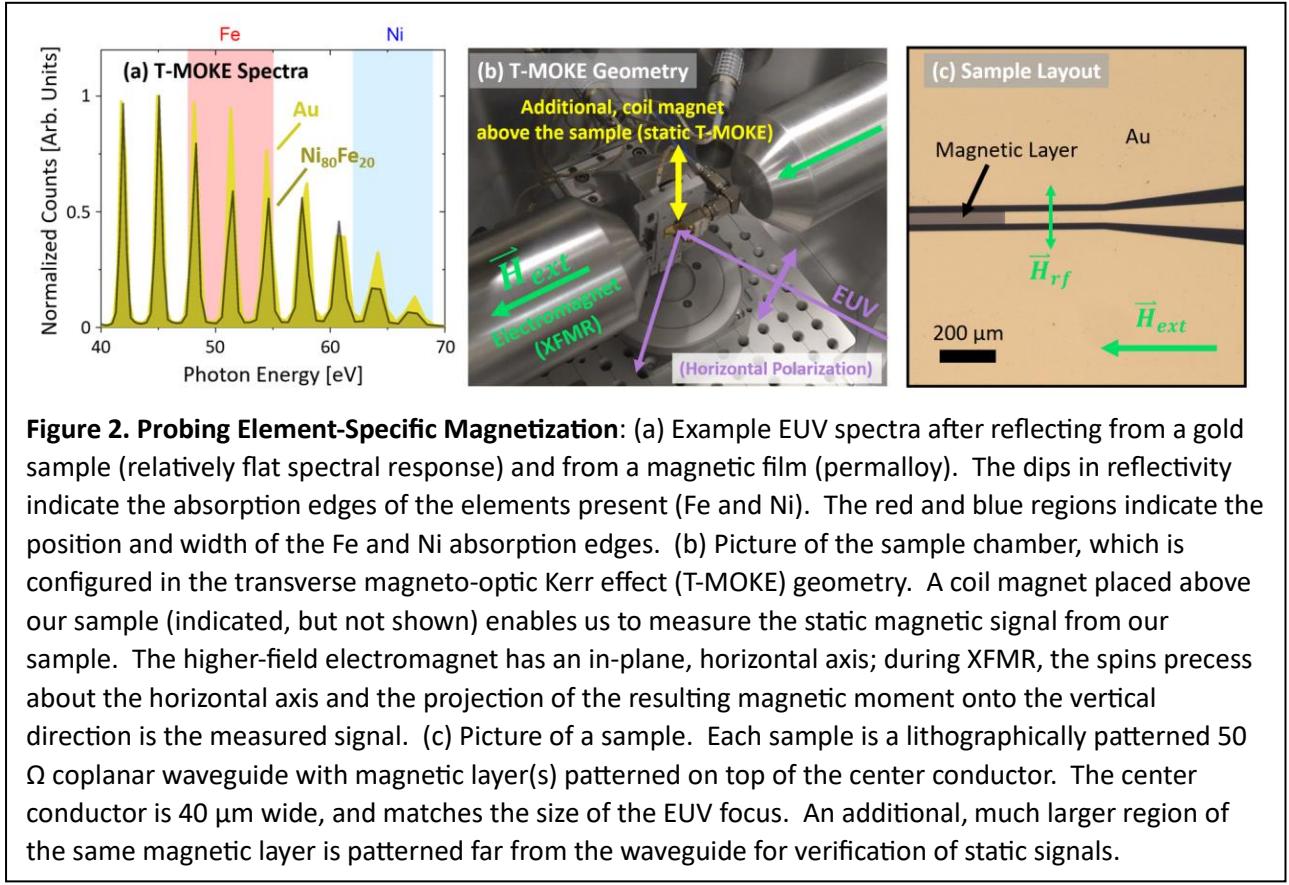

**Figure 2. Probing Element-Specific Magnetization**: (a) Example EUV spectra after reflecting from a gold sample (relatively flat spectral response) and from a magnetic film (permalloy). The dips in reflectivity indicate the absorption edges of the elements present (Fe and Ni). The red and blue regions indicate the position and width of the Fe and Ni absorption edges. (b) Picture of the sample chamber, which is configured in the transverse magneto-optic Kerr effect (T-MOKE) geometry. A coil magnet placed above our sample (indicated, but not shown) enables us to measure the static magnetic signal from our sample. The higher-field electromagnet has an in-plane, horizontal axis; during XFMR, the spins precess about the horizontal axis and the projection of the resulting magnetic moment onto the vertical direction is the measured signal. (c) Picture of a sample. Each sample is a lithographically patterned 50 Ω coplanar waveguide with magnetic layer(s) patterned on top of the center conductor. The center conductor is 40 µm wide, and matches the size of the EUV focus. An additional, much larger region of the same magnetic layer is patterned far from the waveguide for verification of static signals.

Once the EUV is generated, we attenuate the driving IR using two mirrors oriented near Brewster's angle for the IR, and further extinguish the IR by absorbing it in thin metal foil(s) (usually 0.5 µm Al). Then, the EUV passes through a variable-diameter aperture, and is focused onto the sample using a 6° angle of incidence (AOI), grazing-incidence toroidal mirror. This mirror is placed 2 m from the glass capillary and has a focal length of 750 mm, resulting in a spot size of approximately 40 µm diameter [the width of the center conductor of the coplanar waveguide samples, Fig 2(c)]. Since we are using the T-MOKE geometry [see Fig. 2(b)], the sample is oriented near Brewster's angle, specifically at 50° from grazing since this angle maximizes signal from many of our samples (see Fig. 4). We can energize a vertically-oriented coil magnet for static T-MOKE measurements or a horizontally-oriented electromagnet for our XFMR studies. Both magnetic fields are parallel to the sample plane. By reversing the direction of the magnetic field or (for XFMR) by toggling the RF on/off with a fixed magnetic field, we can observe the element-specific magnetic state of the sample as the variation of the spectral peak intensity at the absorption edge of each element. In both cases, the projection of the magnetic moment onto the vertical direction is the detected magnetic signal — for XFMR, the signal will oscillate sinusoidally with the spins' precession. Finally, the reflected EUV strikes a 400 mm toroid in a 4f imaging configuration (magnification of 1) and a 500 grooves/mm blazed grating (period: 2 µm) oriented in the conical geometry, and is collected on a scientific EUV CCD detector. This results in a spectrum with separated peaks, shown in Fig. 2(a).

**Generation of synchronized >60 GHz microwaves**

The pulse width of the EUV is <30 fs, which sets an intrinsic frequency limitation well above 10 THz. Therefore, the practical frequency limitation for XFMR with this instrument is set by the signal-to-noise ratio (SNR, which is sample-dependent and decreases at higher frequency due to smaller precession cone angles) and the timing jitter between the microwave excitation and the EUV probe, which has a measured standard deviation of roughly 1.2 ps, as seen in Fig. 3(a).

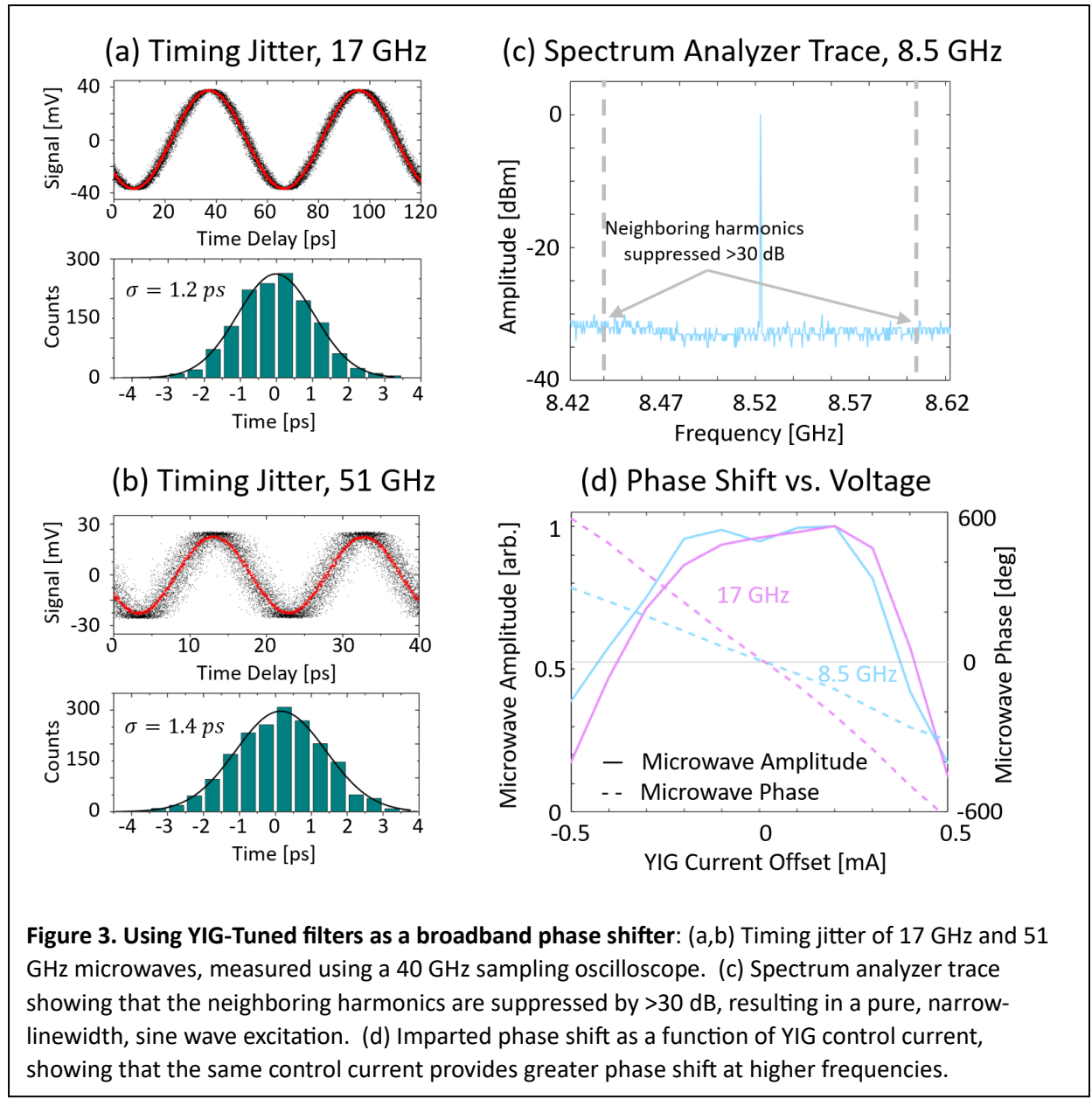

**Figure 3. Using YIG-Tuned filters as a broadband phase shifter**: (a,b) Timing jitter of 17 GHz and 51 GHz microwaves, measured using a 40 GHz sampling oscilloscope. (c) Spectrum analyzer trace showing that the neighboring harmonics are suppressed by >30 dB, resulting in a pure, narrow-linewidth, sine wave excitation. (d) Imparted phase shift as a function of YIG control current, showing that the same control current provides greater phase shift at higher frequencies.

To minimize the timing jitter between the microwave excitation and the EUV probe, we generate the microwaves directly from the IR laser. In particular, we use a similar approach to that taken at synchrotron facilities, wherein an RF frequency comb generator (FCG) produces harmonics of the master clock that times the electron bunches.[9,19] In this case, we adopt the Ti-sapphire oscillator itself as the master clock (81.6 MHz). A portion of the oscillator beam is directed onto a photodiode, and the resulting electrical signal drives an RF frequency comb generator that produces harmonics up to at least 18 GHz (> 220th harmonic). Since the oscillator is simultaneously used to seed the regenerative amplifier, the EUV pulses are intrinsically synchronized to the microwave radiation.

An upper bound of the short-term phase jitter can be made by measuring the microwave signal with a sampling oscilloscope that is triggered on a fast photodiode (rise time: 30 ps) that views the 3 kHz laser that drives the HHG process. Shown in Fig. 3(b) is a measurement of 51 GHz microwaves made in this way, which shows a timing jitter of 1.4 ps. Due to the limited bandwidth of the available microwave components, we used different configurations for 8.5 GHz, 17 GHz, and 62 GHz frequencies (see Fig. S2). Importantly, the timing jitter was roughly the same (<1.4 ps) for each configuration, even when we used additional components like frequency multipliers. The actual jitter is lower than 1.4 ps, owing to the fact that smaller values are beyond the measurement bandwidth of the oscilloscopes. In fact, we determined the intrinsic jitter of the sampling oscilloscope to be 0.8 ps. After removing this contribution from the total jitter (1.2 – 1.4 ps), the upper limit on jitter between the comb generator and the laser amplifier is estimated as 1.0 – 1.1 ps. Furthermore, the timing jitter may be further improved by use of a high-speed photodiode as input to the FCG, instead of the current one which has 1 ns rise time. Considering this upper limit of timing jitter, there is no fundamental limitation to achieve measurement frequencies of 100 GHz or higher.

While we need to excite FMR with a single frequency, the FCG outputs a large series of spectral lines that span several hundred harmonics of the fundamental 82 MHz repetition rate of the oscillator. As shown in the simplified schematic diagram in Fig. 1(b) and in the more complete diagrams in Appendix B, we use a pair of YIG filters as a narrow-band filter to isolate a single frequency or "tooth" of the frequency comb [suppressing other frequencies by more than 30 dB, as shown in Fig. 3(b)]. In addition to filtering out undesired frequencies, the YIG filters shift the phase of the signal as they are detuned from their resonance; we exploit this property for use as a self-contained broadband microwave phase shifter that requires no additional components (see Appendix A). Of course, the total phase range is limited by the width of the YIG resonance. Using two YIGs in tandem (instead of one) gives a larger range of accessible phase shifts while keeping the amplitude relatively flat. For the frequencies investigated here, using two YIG filters enables the phase to be continuously shifted over a range of at least 270 degrees (8.5 GHz), 540 degrees (17 GHz), and with the range generally increasing in proportion to frequency [see Fig. 3(d)].

With our current setup, we are able to vary the base frequency of the FCG from 8.5 GHz to 12 GHz. This limitation is simply set by the bandwidth of the microwave components (including YIG filters, amplifiers, circulators, high-speed switch, etc.) that we had available on hand; however, standard, commercially available products can be used for other/wider bands -- extending both to lower and higher frequency ranges. In our current setup, higher frequencies are achieved by inserting frequency multipliers (2x and 3x) that enable us to extend the output frequencies to values exceeding 60 GHz (see Appendix B). The advantage to this approach is that most of the circuit remains unchanged (i.e., operates at its base bandwidth) since the frequency multiplier is inserted just before the sample, requiring only one higher frequency amplifier and components after the multiplier. Figure 3(a, b) shows an example of a 17 GHz signal generated by frequency doubling 8.5 GHz, and a 51 GHz signal generated by using both a frequency doubler and a frequency tripler. A final note is that, by introducing a circulator and a lock-in amplifier [shown in Fig. 1(b)], we are able to perform *in situ* inductive FMR measurements on the sample. This not only provides information about the optimal fields to use during the XFMR measurement, but also allows us to monitor the sample for any possible damage, heating, or other changes during the measurement.

**XFMR Procedure**

After we have generated femtosecond-duration EUV light in the appropriate range of photon energies (45–70 eV) and generated microwaves that are phase-synchronized to those EUV pulses, we can perform XFMR using the following steps. We first set the excitation microwave frequency (e.g., to 17 GHz) by coarsely tuning the YIG filter currents to select a tooth of the RF frequency comb. Fine-tuning the currents about this point will control the phase shift of the microwaves. To quantify this *in situ*, we use a sampling oscilloscope, which is triggered on the 800 nm laser that drives HHG, to measure the microwaves, and sweep the YIG voltages independently to determine the phase shift per volt (approx. 600 deg/mA at 8.5 GHz) and the accessible range of phase shifts.

After setting the frequency and calibrating the phase-shifting YIG control currents, we use our *in situ* inductive FMR setup to determine the magnetic fields to perform XFMR (i.e., resonance field and linewidth), as well as the proper microwave power to maximize our signal but suffer no effects from heating and ensure that we stay within the linear regime of the magnetization dynamics. The inductive FMR trace will show a linear combination of the real and imaginary parts of the microwave reflected from the sample. It is easiest to interpret the imaginary part, which we achieve by use of RF phase shifting components.

Once we know the fields over which to perform XFMR, we send EUV light onto the sample. We can then measure and optimize our system using a static T-MOKE spectrum obtained by energizing a small coil magnet above our sample. By switching the polarity of the current, the magnetic state of the sample will reverse and this will write itself onto the harmonic spectrum; in other words, the harmonic peaks at the photon energies within the Fe M-edge will change intensity proportionally to the magnetic moment of the Fe. By measuring the distance between the grating and the camera and then counting the number of pixels between the specular beam and the harmonic peaks of interest, we can estimate the photon energy of each harmonic peak and determine which element is magnetized by using the grating equation ($\lambda = T \sin(\theta) / q$, where $\lambda$ is the wavelength, $T$ is the period of the grating, $\theta$ is the diffraction angle, and $q$ is the diffraction order). To improve the SNR of our static T-MOKE measurements, we average several frames (typically 20 frames per magnetic field polarity) and generally use an exposure time of 5 seconds for each frame while 4x4 binning. To avoid condensation issues on the camera sensor due to operating in only high vacuum conditions, we do not cool the sensor.

Finally, if we fix the applied magnetic field and the YIG currents, then each EUV pulse will impinge on the sample at a consistent phase of the applied microwaves, and therefore at the same precessional phase of the spins within the sample. We again use 5 second exposure times for each frame and average over many frames (typically 100 with microwaves applied and 100 without). A difference spectrum is obtained by recording alternating spectra with the microwaves turned on/off and subtracting them from each other. As a result, the difference spectrum is proportional to the projection of the magnetization onto the vertical axis at a particular phase in the precession; as the phase is shifted through 360°, this measures a sinusoidal response. This process can be repeated at a series of magnetic fields to map out the full, element-resolved, field-swept ferromagnetic resonance peak.

To maximize the SNR, we adopt an on/off measurement scheme, in which every other measured spectrum is made without the microwave signal applied. By subtracting the spectrum with the microwaves off from that with the microwaves on, the effect of longer-timescale fluctuations of the EUV intensity can be minimized. Since our exposure times were 5 seconds, the effects of fluctuations on this timescale or shorter are not suppressed. We partially correct these shorter-term instabilities by assuming that they are perfectly correlated across all the harmonic peaks (including those with no magnetic signal) and dividing all peaks by the frame-to-frame fluctuations of the peak intensities far away from the magnetic absorption edges. This is only approximately correct, but consistently improves our data quality.

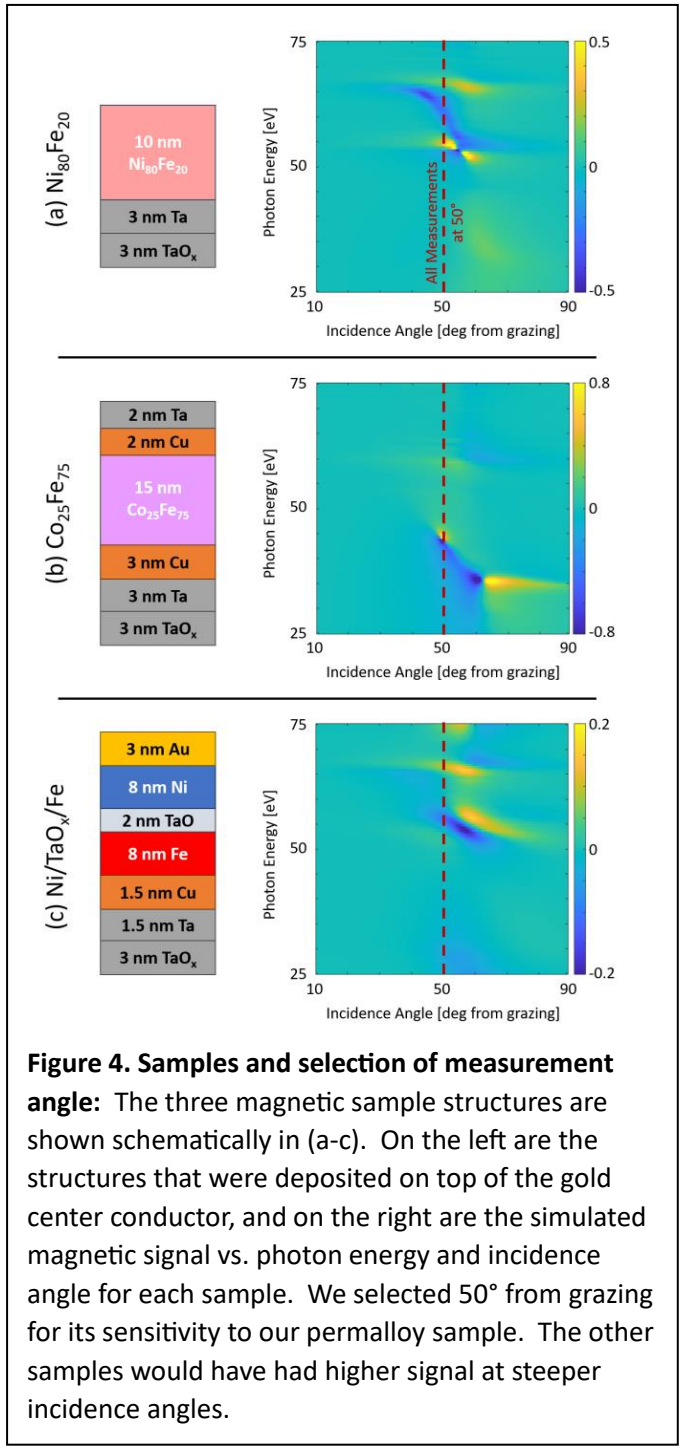

**Figure 4. Samples and selection of measurement angle:** The three magnetic sample structures are shown schematically in (a-c). On the left are the structures that were deposited on top of the gold center conductor, and on the right are the simulated magnetic signal vs. photon energy and incidence angle for each sample. We selected 50° from grazing for its sensitivity to our permalloy sample. The other samples would have had higher signal at steeper incidence angles.

**Sample Fabrication**

Samples were fabricated using optical lithography combined with lift-off processes on thermally oxidized Si wafers. The first step was to pattern the waveguide structure. This consists of a coplanar waveguide (CPW) structure that is designed to mate with a commercial high-frequency end-launcher on one end.

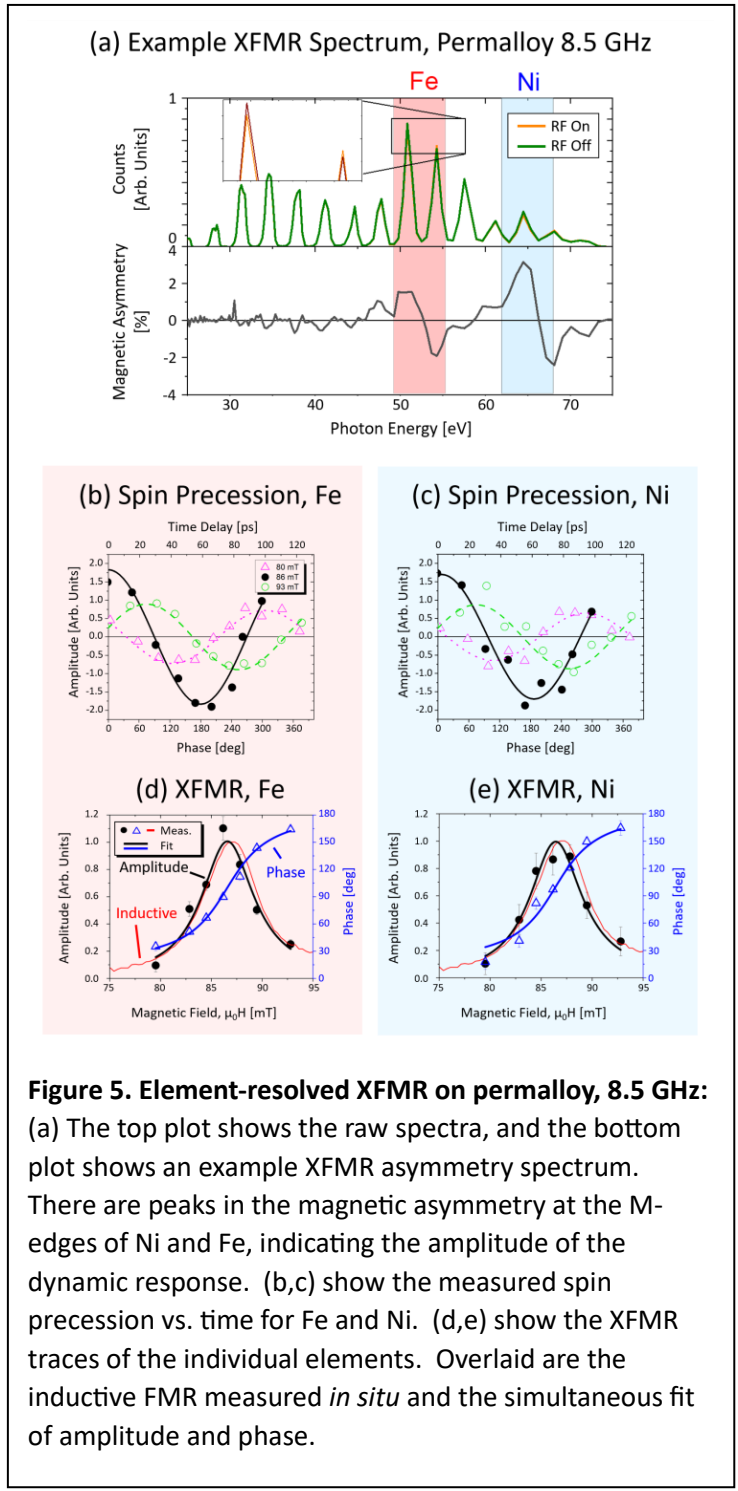

**Figure 5. Element-resolved XFMR on permalloy, 8.5 GHz:** (a) The top plot shows the raw spectra, and the bottom plot shows an example XFMR asymmetry spectrum. There are peaks in the magnetic asymmetry at the M-edges of Ni and Fe, indicating the amplitude of the dynamic response. (b,c) show the measured spin precession vs. time for Fe and Ni. (d,e) show the XFMR traces of the individual elements. Overlaid are the inductive FMR measured *in situ* and the simultaneous fit of amplitude and phase.

This connection is designed to maintain a 50 Ohm impedance, minimizing reflections or losses across that transition to/from the end launcher. The CPW then tapers down to a center conductor width of 40 μm in most cases, but other widths are also used. To form the CPW, 5 nm Ti and 100 nm of Au was evaporated and lifted off. A second lithography step was used to define the magnetic layer of interest. This was patterned on top of the narrow section of the CPW, as shown in Figure 2(c). The magnetic material was then deposited using magnetron sputtering. In all cases, a 3 nm layer of $TaO_x$ was first deposited prior to depositing the magnetic layer in order to prevent electrical contact between the magnetic layer and the CPW. This also prevents any spin-pumping losses when the magnetic layer is driven to FMR. After lift-off of the magnetic layer, the wafer was diced using a diamond saw.

In this study, we focus on three magnetic samples, whose structures are summarized in Figure 4. First, we performed XFMR on permalloy at 8.5 GHz. This sample could not be used to demonstrate XFMR at frequencies above 11 GHz due to the magnetic field limitation of our current setup (max $\mu_0 H$ = 180 mT). We instead focus on a $Co_{25}Fe_{75}$ sample to achieve higher-frequency XFMR measurements at 17 GHz. The increased saturation magnetization of $\mu_0 M_s$ = 2.4 T increases the FMR frequency to more than 17 GHz for the magnetic fields that we have available.[56] Finally, to show the ability to measure independently the dynamics of different layers within a multilayer sample, we measured a $Ni/TaO_x/Fe$ multilayer.

**XFMR Results: Permalloy at 8.5 GHz**

As a first demonstration, we measured the XFMR response of a 10 nm thick $Ni_{80}Fe_{20}$ (permalloy) sample. The individual spectra at the microwave phase giving the maximum magnetic signal are shown in Figure 5(a). Two spectra are shown: one each with the microwaves turned on and turned off. There is a small but measurable difference between the two spectra that measures spin precession within the sample; this signal is significantly enhanced at the expected photon energies corresponding to the M-edges of Ni and Fe, and varies sinusoidally with the phase of the applied microwave excitation. The difference between the two curves is related to the magnetic asymmetry, which is calculated as $(I_{on} - I_{off})/(I_{on} + I_{off})$, or the difference between the intensities of the spectra when the rf field is on $I_{on}$ and when it is off, $I_{off}$ divided by the sum of the intensities. The magnetic asymmetry is shown in the bottom half of Figure 5(a). For the sake of visualization, we suppress the noise in the asymmetry spectra in the regions between the harmonics where there is little to no signal; we do so by adding a large value to the denominator when the intensity is below a threshold that is set to a value just above the minimum counts in between the harmonics. This minimizes the effect of amplifying noise in the spectral region where there is little to no signal and/or prevent a divide by zero. It is important to note that this is not used in the quantitative fitting of the spectra, but solely to improve visualization.

The amplitude of the asymmetry should vary sinusoidally at the microwave driving frequency. To capture the amplitude and phase of the response for each element, the magnetic spectra are recorded at several values of the rf phase. The amplitudes at the Fe and Ni M-edges are plotted as a function of the phase in Figures 5(b) and 5(c), respectively. A sinusoid is then fit to the data, using the frequency measured by a spectrum analyzer during the experiment. This process is repeated for all values of the applied bias field — a few examples are also shown in Figures 5(b) and 5(c). It is clear that, as expected, both the amplitude and phase of these sinusoidal behaviors change as the magnetic field is also changed. Figure 5(d) and 5(e) are element-resolved XFMR plots. That is to say, they show the amplitude (black circle) and phase (blue open triangle) obtained from the sinusoidal fits of the Fe and Ni peaks data for several values of the applied magnetic bias field. The amplitude $A$ and phase $\varphi$ are simultaneously fit to the imaginary component dynamic susceptibility Im[χ] and phase $\varphi$ derived from the Landau-Lifshitz equation as defined by the following equations[7]:

$$Im[\chi(f)] = A_0 H_1 \left[\frac{-H_3}{H_2^2 + H_3^2}\right]$$

$$Re[\chi(f)] = A_0 H_1 \left[\frac{H_2}{H_2^2 + H_3^2}\right]$$

$$\varphi = atan\left[\frac{Re[\chi(f)]}{Im[\chi(f)]}\right] + offset$$

where, $H_1 = H + M_{eff}$, $H_2 = HH_1 + H^2_{eff}$, $H_3 = \Delta H(2H+M_{eff})$, $H_{eff} = 2\pi f/\gamma\mu_0$ is the effective field, $\gamma$ is the gyromagnetic ratio, $\mu_0$ is the permeability of free space, $f$ is the frequency, $H$ is the applied field, $M_{eff}$ is the effective magnetization, $\Delta H$ is the linewidth, and $A_0$ is an amplitude. Unique fits for $M_{eff}$ and $H_{eff}$ are not possible when fitting to a single resonance curve, so we set $\mu_0 M_{eff}$ equal to 1 T for the fits. Since the goal of this work is not to extract values for these parameters, such an assumption was used solely to allow the fit to converge. The simultaneous fit is indicated in the figure as solid lines. The identical response between the Fe and Ni is expected based on the strong exchange coupling and standard models for FMR. This provides rudimentary validation of the approach. Furthermore, the quality of the simultaneous fit highlights the agreement between the measured and expected behavior as the external bias field is swept through the ferromagnetic resonance. To confirm the measurement of the FMR, we overlay as a red line the field-swept FMR spectrum that we obtain via the *in situ* inductive method by measuring the amplitude of the reflected microwaves from the CPW sample structure. Good agreement is seen between the XFMR and inductive FMR methods. The small difference in linewidth between the inductive FMR and XFMR could be due to the fact that XFMR is probing a localized area of the sample, whereas the inductive approach is probing the entire sample which will likely have more inhomogeneity. This is especially true given the fact that the current sample was fabricated in a lift-off process on a thick Au layer. This produces edge regions with both line edge roughness and thickness variation, coupled with the fact that the Au underlayer itself induces additional film roughness. Finally, we note that since our EUV beam is focused in a roughly-Gaussian spot with a peak intensity near the center of the waveguide, our EUV measurements intrinsically probe the center region of the waveguide more than the edges.

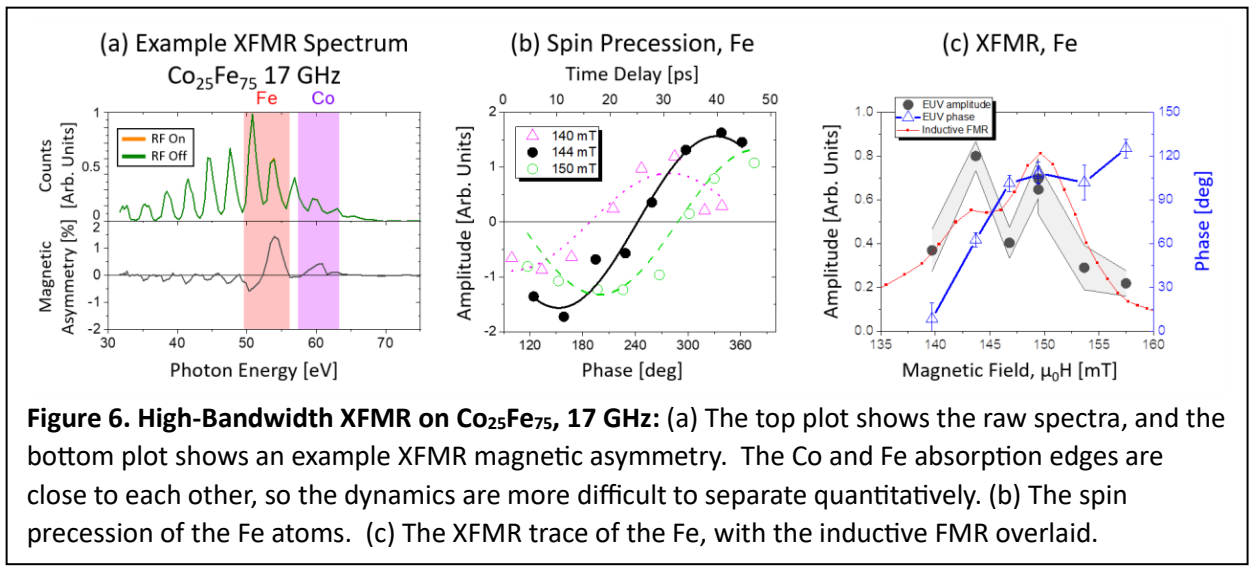

**Figure 6. High-Bandwidth XFMR on $Co_{25}Fe_{75}$, 17 GHz:** (a) The top plot shows the raw spectra, and the bottom plot shows an example XFMR magnetic asymmetry. The Co and Fe absorption edges are close to each other, so the dynamics are more difficult to separate quantitatively. (b) The spin precession of the Fe atoms. (c) The XFMR trace of the Fe, with the inductive FMR overlaid.

**XFMR Results: $Co_{25}Fe_{75}$ at 17 GHz**

The second sample that we studied is $Co_{25}Fe_{75}$; due to its high $M_s$, we were able to measure XFMR at 17 GHz even with magnetic fields as low as $\mu_0 H \approx 150$ mT. Notably, at 50° from grazing, our setup has a low contrast for the magnetic signal at the Co M-edge as shown in Figure 4(b). This limitation is solely due to the current incidence angle of our system, and can be overcome by changing this angle to 52°. Regardless, we are still able to measure the dynamics of each element, as shown in Figure 6 (a). Broadly speaking, the Co and Fe behave the same, as expected; however, without more careful data analysis, quantitative XFMR traces cannot be shown for Co and Fe independently (e.g., to compare phase shifts) due to the proximity of the Co (57–62 eV) and Fe (50–55 eV) M-edges.

The inductive FMR trace shown in Figure 6(c) shows a double-resonance, which we suspect is due to inhomogeneity in the sample. This material is known to have large inhomogeneity when deposited on rough surfaces or when non-ideal seed and capping layers are used.[57] The XFMR trace also shows this double peak structure, and the locations of the two resonances agree in both methods. As expected, the phase in the XFMR is not a clean arctan due to superposition of the two resonances. The relative amplitudes of the two peaks are not the same in the XFMR and inductive FMR, which is likely due to the fact that the EUV probes only a small region (~60 μm diameter), whereas inductive FMR probes the entire structure. We do not think this comes from uncertainty in our XFMR data; in addition to estimating the error bars from the quality of the sinusoidal fits (i.e., the variance in the data), we tested the repeatability of the XFMR measurement by measuring at 150 mT both at the beginning and end of the measurement [included in Fig. 6(c)]. Within their respective error bars, these measurements at 150 mT agree and demonstrate the stability and robustness of the measurement system.

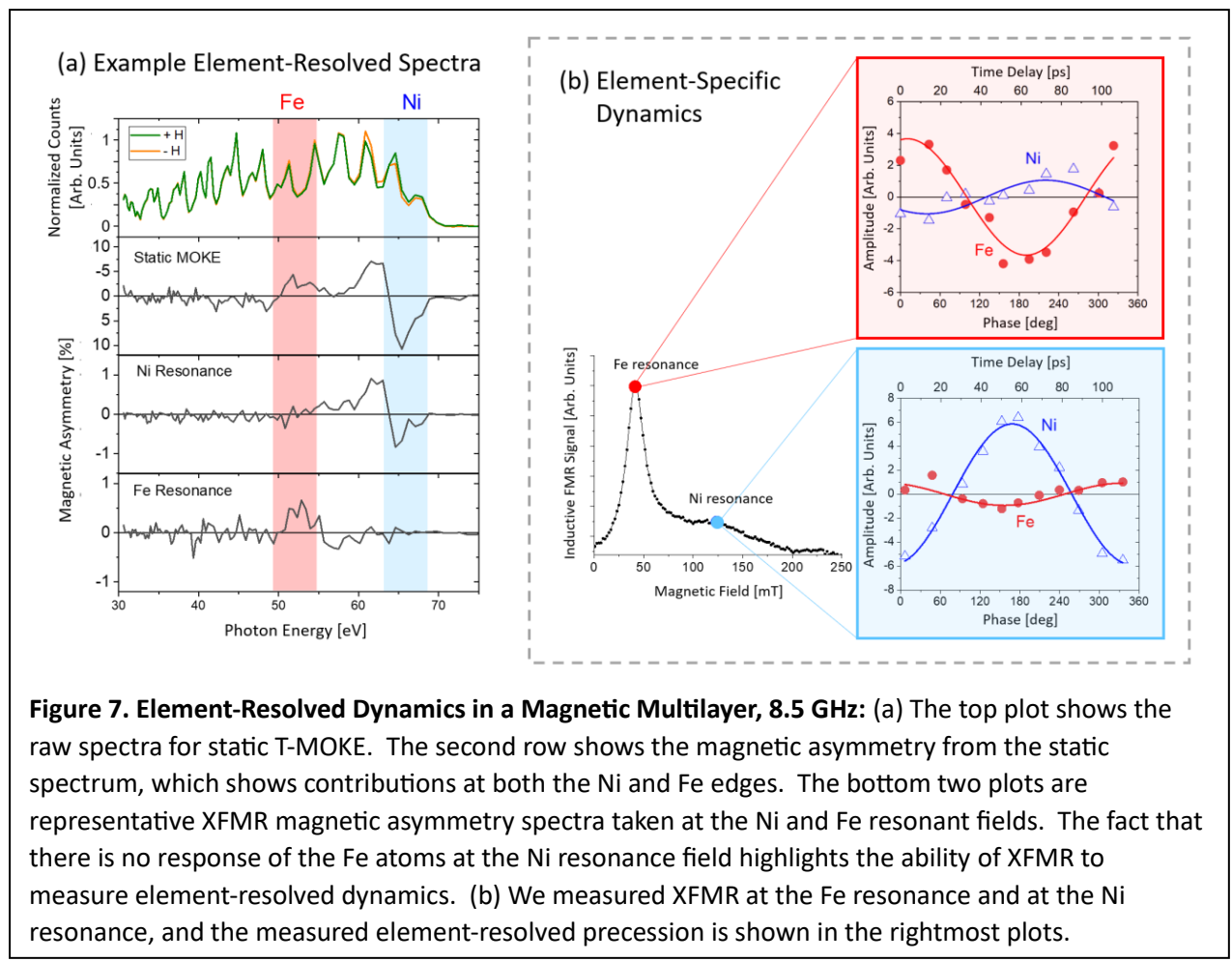

**Figure 7. Element-Resolved Dynamics in a Magnetic Multilayer, 8.5 GHz:** (a) The top plot shows the raw spectra for static T-MOKE. The second row shows the magnetic asymmetry from the static spectrum, which shows contributions at both the Ni and Fe edges. The bottom two plots are representative XFMR magnetic asymmetry spectra taken at the Ni and Fe resonant fields. The fact that there is no response of the Fe atoms at the Ni resonance field highlights the ability of XFMR to measure element-resolved dynamics. (b) We measured XFMR at the Fe resonance and at the Ni resonance, and the measured element-resolved precession is shown in the rightmost plots.

**XFMR Results: Ni/Fe Multilayer at 8.5 GHz, showing element-specific dynamics**

The third and final sample in this study is a multilayer sample with Ni and Fe layers that are separated by an insulating 3 nm $TaO_x$ layer. The magnetic isolation of the Fe and Ni layers should yield independent dynamics of the layer except for some potential dipolar coupling. Figure 7(b) shows the *in situ* inductive FMR response of the sample taken at 8.5 GHz, wherein two peaks are observed. The high amplitude and narrow peak near 40 mT is presumably due to the Fe layer since Fe has higher $M_s$ and lower damping relative to Ni.[58] The broad and low amplitude peak near 120 mT is likely from the Ni, by the same argument. Indeed, XFMR measurements at 40 mT reveal that the precession amplitude of the Fe layer is significantly higher than that of the Ni. Given the broad linewidth of the Ni peak and the potential for dipolar coupling between the layers, it is not surprising to see a non-zero precession angle for Ni as well. Repeating this measurement at $\mu_0 H$ ~125 mT shows the opposite behavior, where the precession amplitude of the Ni signal is significantly more than that of the Fe. Furthermore, in both cases there is a noticeable phase shift between the precession of the two elements. Such a phase shift is to be expected since the resonance fields (and therefore phase shift through the resonance) are at different locations. Any dipolar coupling would also induce a relative phase shift. Measurements taken at more fields across the resonances would in principle enable us to determine the amount and type of coupling between the layers. Measuring such phase shifts in structures where the coupling is more complicated (e.g., from spin currents or weak exchange) allows quantification of such interactions. This highlights the ability to separate dynamics within different layers or sublattices in scientifically interesting or industry-relevant samples.

**Discussion and Outlook**

This work shows that high-bandwidth, element-specific measurements of dynamic magnetic phenomena can be achieved in a laboratory. Despite the fact that no reference beam was used to correct for intensity fluctuations and drift, small deviations of the magnetization with cone angles below 5 degrees were easily observed at high-frequency. Going forward, many straightforward improvements can be made to increase the sensitivity and SNR, allowing more subtle phenomena to be observed along with significantly smaller cones angles and allow for more field steps across the resonance. The latter will be particularly important as measurements are pushed to higher frequencies. Here, we discuss several improvements that can be made to greatly improve performance.

First, a reference beam that provides reference spectra of the EUV light incident on the sample can be used to normalize and divide out intensity fluctuations on both short and long timescales. The absence of this in the current setup represents one of our biggest barriers to achieving higher SNR. The frame-to-frame RMS intensity is approximately 5% for 5 sec exposures, and there is also similar drift on longer time scales. Furthermore, the fluctuations of each harmonic are not necessarily correlated (though our analysis assumes that they are). Such normalization is critical at synchrotrons since the beam intensity also varies significantly over time. Recently, the use of a reference beam in HHG systems used to measure spin dynamics was shown to increase SNR by almost an order of magnitude.[59–61]

Second, reflection-geometry measurements as shown here can be significantly improved by incorporating measurements at multiple angles.[39,55] This can be seen in Figure 4, which shows a calculation of the T-MOKE contrast for the samples used in this study. This calculation was performed using the scattering matrix approach described in Ref. [62]. While the 50° angle of incidence (measured from grazing incidence) used here is near the maximum-contrast point for the $Ni_{80}Fe_{20}$ sample, we would achieve better SNR and better-separated Ni and Fe peaks in the $Co_{25}Fe_{75}$ and Ni/Fe multilayer samples if the scattering angle was changed to 52 degrees. Incorporating the ability to change the incidence angle easily will enable us to optimize the signal for general samples, as well as improve the depth-sensitivity of our measurement.

Third, while we have shown adequate synchronization of the EUV pulses to 62 GHz microwave fields, the timing jitter we measure supports extension to significantly higher frequencies by using commercially available components. Furthermore, it may be possible to improve the timing jitter by replacing the photodiode that is used for the input to the frequency comb generator. We found that using a sharper rise time on the input to the frequency comb generator reduced the timing jitter. Since the current photodiode still has a relatively long rise time (low bandwidth) of 1 ns, which depending on the electronic noise present can produce additional timing jitter, we anticipate that using a photodiode with a shorter rise time will further improve the timing jitter. Also, in addition to using multipliers, synchronized 100+ GHz tones with higher signal-to-noise ratio can be generated by driving an electrooptic modulator with a 10 or 20 GHz harmonic of the 80 MHz laser repetion rate.[63] We further note that our current Ta-sapphire oscillator is passive and lacks any active frequency stabilization. Femtosecond-level timing jitter between optical and electronic signals is accessible with techniques that employ stabilization of the underlying Ti:sapphire frequency comb mode spacing and carrier-envelope offset frequency.[63–65]

Finally, as already mentioned, this technique will be extended to include an imaging modality whereby the coherence of the EUV will be exploited.[51,53] Lensless imaging techniques such as coherent diffractive imaging, ptychography, and holography have been used with EUV and x-ray light to image nanostructures and magnetic textures with resolution smaller than the illumination spot size and approaching the diffraction limit.[66,67] Furthermore, these techniques provide quantitative measurement of the phase shift imparted by the sample, which provides a dramatic improvement in contrast to certain sample features, such as topography and oxidation state.[55] Combining such a technique with the XFMR instrument developed in this paper will enable us to perform dynamic, *in situ* measurements of functioning devices and elucidate the underlying physics.

**Summary**

We have demonstrated the ability to bring element-specific, x-ray detected ferromagnetic resonance spectroscopy (XFMR) to a laboratory setting. We did so by combining a high-harmonic generation (HHG) light source, which generates 35–70 eV light, with an RF frequency comb generator, and were able to generate phase-stable microwaves (timing jitter < 1.1 ps) above 60 GHz that we can use for XFMR measurements at the M-edge of most magnetic elements. The instrument can perform measurements in transmission or reflection; we showed reflection-mode measurements here that use the transverse magneto-optic Kerr effect (T-MOKE) geometry for magnetic contrast even on samples with opaque substrates (e.g., thick silicon). We used our instrument to perform high-frequency XFMR measurements on three samples: permalloy (8.5 GHz), $Co_{25}Fe_{75}$ (17 GHz), and a Ni/Fe multilayer sample that showed different dynamics in the Ni and Fe layers (8.5 GHz).

This system provides the capability in the near future to measure element-, layer-, or sublattice-specific dynamics in industry-relevant and scientifically interesting magnetic thin films. Moreover, by introducing variable-angle measurements into this system, we will enhance the SNR and depth-sensitivity of the technique to be able to monitor precisely the spin transport and accumulation across interfaces. Finally, the spatial coherence of the source, combined with the ability to synchronize electrical pulses to the EUV probe, lays the foundation for performing these measurements in-operando on individual or arrays of functional spintronic devices.

**Appendix A: Tuning the YIG Pair**

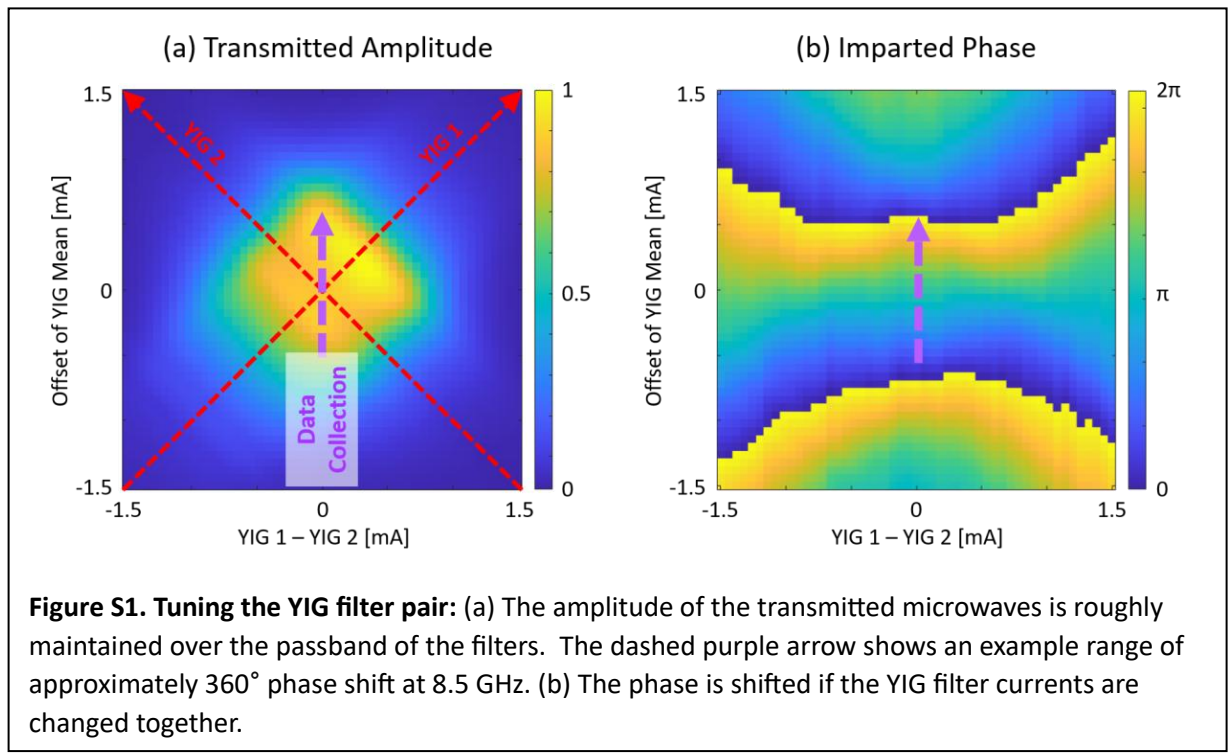

**Figure S1. Tuning the YIG filter pair:** (a) The amplitude of the transmitted microwaves is roughly maintained over the passband of the filters. The dashed purple arrow shows an example range of approximately 360° phase shift at 8.5 GHz. (b) The phase is shifted if the YIG filter currents are changed together.

We are able to tune the pair of YIG filters by using a pair of part-per-million (PPM)-stable current sources (one per YIG). In particular, we drive the YIGs with roughly 200 mA, and changing the current by 1 μA shifts the phase by roughly a degree at 8.5 GHz [shown both in Fig. 3(d) and Fig. S(1a)]. One YIG alone does not provide 360° phase shift at 8.5 GHz (or below) before the transmitted amplitude starts to decline. By using two YIGs and sweeping the currents together, we are able to shift the microwave phase over a full cycle while maintaining a relatively constant microwave amplitude. Fig. S1 shows the amplitude and phase of the transmitted microwaves as the YIG filters are tuned together (vertical) and apart (horizontal).

**Appendix B: Complete RF Generator Schematic**

Figure S2 shows the full schematic for the microwave generator that was used to generate 8.5–62 GHz. We use frequency multipliers to achieve frequencies above 13 GHz. Above 54 GHz, we encounter significant losses in our system since we exceed the bandwidth of several components, as well as the 40 GHz sampling oscilloscope. Despite this fact, we are still able to evaluate the timing stability and test the synchronization of microwaves to the EUV pulses up to 62 GHz. Figure S2 shows that we can still measure the synchronization our EUV pulses up to frequencies of 62 GHz, albeit with increased noise due to microwave losses. Additionally, the final bandpass filter fails to suppress the 41 GHz harmonic also output by the tripler when generating 62 GHz microwaves, leading to some beating of the signal. However, these are not fundamental limitations since higher bandwidth components are commercially available.

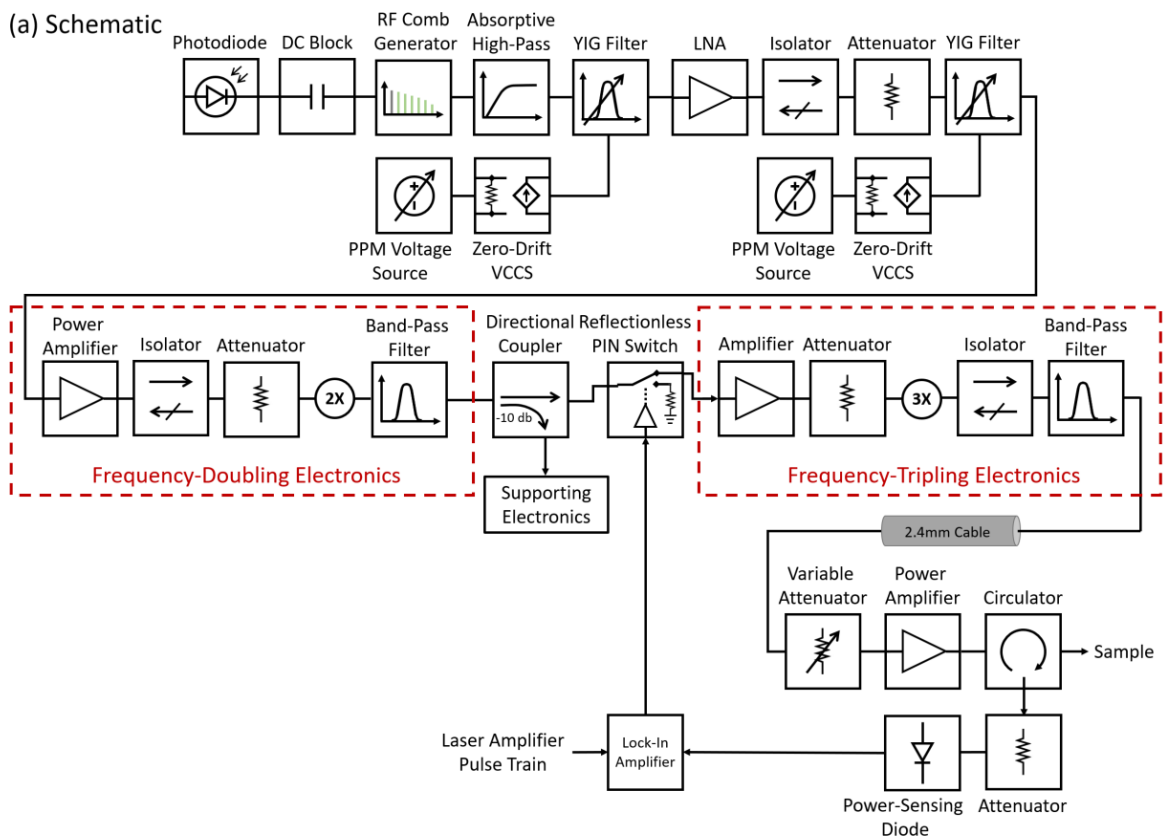

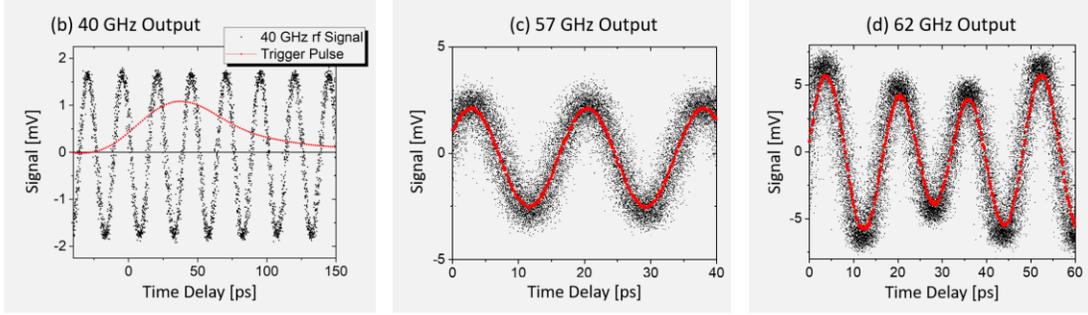

**Figure S2. Complete RF Generation Circuit Schematic for up to 62 GHz:** (a) The full circuit schematic for generating synchronized microwaves is shown. The frequency-doubling and frequency-tripling electronics are omitted when generating lower frequencies. (b-d) Traces measured with our 40 GHz bandwidth sampling scope when the circuit is generating 40 GHz, 57 GHz, and 62 GHz microwaves, respectively. In (b), the trigger pulse (measured by a high-speed, real-time oscilloscope) is overlaid. Note that the beating at 62 GHz is due to exceeding the rated bandwidth of multiple components, most notably the final bandpass filter passes 41 GHz in addition to the desired 62 GHz.

**Acknowledgments**

The authors are grateful to Henry Kapteyn, Margaret Murnane, and Tom Silva for valuable discussions and advice, and to Dmitriy Zusin and Christian Gentry for developing the multilayer reflectivity calculating code.  MT and JW acknowledge funding support from the National Research Council (NRC) Post-Doctoral Fellowship program.  HTN acknowledges support through the NIST cooperative agreement 70NANB18H006 with the University of Colorado Boulder.